\documentclass[aps,prl,twocolumn,reprint,10pt,amsmath,amssymb,floatfix,showpacs]{revtex4-1}

\usepackage{graphicx}

\def\SdotS{\mathbf{S}_i\cdot\mathbf{S}_j}
\newcommand{\comment}[1]{}

\begin{document}

\title{Orbital order, stacking defects and spin-fluctuations in the $p$-electron molecular solid RbO$_2$}
\author{E. R. Ylvisaker}
\author{R. R. P. Singh}
\author{W. E. Pickett}
\affiliation{Department of Physics, University of California, Davis, California, 95616}

\date{\today}

%
\begin{abstract}
We examine magnon and orbiton behavior in localized O$_2$ anti-bonding molecular 
$\pi^*$ orbitals using an effective Kugel-Khomskii Hamiltonian derived from a two band Hubbard model with 
hopping parameters taken from {\em ab initio} density functional calculations.  The considerable difference 
between intraband and interband hoppings leads to a strong coupling between the spin wave dispersion and 
the orbital ground state, providing a straightforward way of experimentally 
determining the orbital ground state from the measured magnon dispersion.  The near degeneracy of different 
orbital ordered states leads to stacking defects which further modulate spin-fluctuation spectra.  Proliferaion 
of orbital domains disrupts long-range magnetic order, thus causing a significant reduction in the observed 
N\'eel temperature.
\end{abstract}



\maketitle
 
Correlated systems have generated considerable interest in the literature in recent years.  The discovery of high
temperature superconductors and the subsequent development and application of correlated methods like LDA+U
\cite{LDAUreview} and DMFT \cite{RevModPhys.68.13}, has led to remarkable success in dealing with strongly correlated
systems.  At integer filling, strongly correlated systems are typically insulating and often show antiferromagnetic
behavior arising from exchange or superexchange processes.  Such systems include the undoped
cuprates, where there is one hole per site that can hop in a square lattice of 
Cu $d_{x^2-y^2}$ orbitals, and heavy fermion materials like CeCuIn$_5$ where the Ce $4f$ 
orbitals weakly couple to the valence states.  Multiband correlated systems can also show orbital ordering; the 
earliest successful application of LDA+U found orbital ordering in the KCuF$_3$ system \cite{Medvedeva02}.

Since correlated behavior is typically the domain of materials with $3d$ and $4f$ orbitals, comparatively little attention
has been given to the study of correlated behavior in $p$ orbital systems.  However, local moment magnetism in $2p$
orbitals has been implicated in several systems, such as at polar oxide
vacancies \cite{Elfimov02} and substitutionals \cite{Pardo08}.  The occurrence of $2p$ orbital moments at
$p$-type LaAlO$_3$/SrTiO$_3$ interfaces \cite{Pentcheva06} is still the only
viable explanation of the {\it insulating} character that
is observed in these interfaces, where the electron count would suggest metallic interface states.  Alkali hyperoxides, to
be discussed below, comprise another likely example.  Recent calculations \cite{Peng09} suggest that doping of $d^0$ (no
$d$ electrons) magnetic systems can stabilize or even enhance $2p$ magnetic moments in systems such as ZnO nanowires
\cite{Peng09}.

Recently, there has been significant interest in studying correlations in solid molecular systems, such as SrN \cite{Volnianska08}, which consists of Sr octahedra containing either isolated N atoms or N$_2$ dimers, with 
calculations predicting that the magnetic moment is strongly confined to the anionic N$_2^{2-}$ dimers.  Calculations on
the Rb$_4$O$_6$ system \cite{Attema05} and Cs$_4$O$_6$ \cite{Attema07} suggest that  these systems would be a 
half-metallic ferromagnets particularly useful for spintronic applications, due to the reduced spin-orbit interaction 
in $p$ orbitals.  However, these calculations were done within weakly correlated density functional theory; more recent
calculations using LDA+U \cite{Winterlik09} suggest that the valence charge separates to give a mixture of magnetic
hyperoxide O$_2^-$ anions and nonmagnetic peroxide O$_2^{2-}$ anions, and an insulating ground state.  There seems to be
some experimental disagreement as to whether Rb$_4$O$_6$ is conducting \cite{Jansen99} or insulating \cite{Winterlik09}.  
In Rb$_4$O$_6$ the three different orientations of the O$_2$ dimers along the principle axes of the crystal give rise to
frustration of the magnetic order.

Our interest here is in the alkali hyperoxides, taking RbO$_2$
as a specific example.  Solovyev \cite{Solovyev08} has provided a study of the sister compound KO$_2$, considering the
large spin-orbit coupling (SOC) limit.  We present here an alternative viewpoint for RbO$_2$,
based on the supposition that SOC is not so large an effect, so
orbital moments are quenched by the crystal field, making the
conventional real $p_x, p_y$ orbitals the natural basis for studying spin and orbital phenomena in RbO$_2$.  
Experimental measurement of the Land\'{e} $g$-factor \cite{Labhart79} yields values close to 2, indicating a mostly 
spin moment, rather than the $g = 4/3$ value expected for large spin orbit coupling.  

\begin{figure}[t]
\begin{center}
\includegraphics[width=0.45\textwidth]{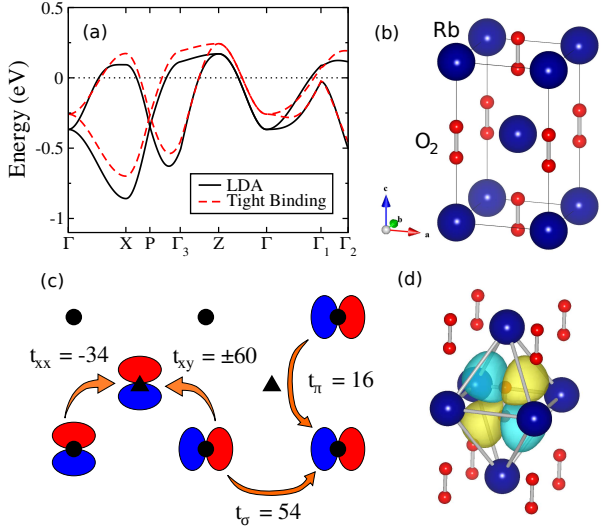}
\caption{(Color online) (a) Paramagnetic band plot of RbO$_2$ showing bands at the Fermi level from both DFT(LDA) and
tight binding.  The bands are filled such that there is one hole per site. (b) Conventional unit cell of the tetragonal
phase of RbO$_2$.  (c) Schematic showing tight binding hopping parameters in meV.  Circles represent O$_2^-$ anions in a
plane perpendicular to the molecular axis, triangles represent O$_2^-$ anions in the nearest planes above or below, Rb not
shown.  (d) Isosurface of the $\pi_x^*$ Wannier function in its local environment.  (b) and (d) produced with Vesta
\cite{vesta}.
\label{fig:tb}}
\end{center}
\end{figure}

The MO$_2$ systems (M=Li,Na,Rb,Cs) exhibit complex phase diagrams and low temperature 
antiferromagnetism \cite{Rosenfeld78}.  The phase diagrams at low temperature consist of several structural changes which
are minor symmetry lowering distortions from the room temperature (averaged) tetragonal phase which is comparable to a
distorted rock salt structure with O$_2^-$ ions playing the
role of the anion, with the molecular axis pointing along the $c$ direction, shown in Fig. \ref{fig:tb}b.  The Jahn-Teller
effect causes the O$_2$ molecules to tilt away from the tetragonal axis, an effect which is difficult to reproduce in a
non-magnetic LDA calculation.  Below 194 K, RbO$_2$ shows incommensurate superstructure and a mixture of pseudotetragonal,
orthorhombic and monoclinic crystal structures that are all slight distortions of the tetragonal phase \cite{Rosenfeld78}.

The standard LDA calculation for RbO$_2$ produces a half-metallic (ferromagnetic)
state in the averaged unit cell \cite{Solovyev08}, or an 
antiferromagnetic metal in a Rb$_2$O$_4$ supercell, in contrast to experimental
reports that MO$_2$ compounds are insulating \cite{Jansen99}.  As seen in Fig. \ref{fig:tb}a, the O$_2$ $\pi^*$ 
bands near the Fermi level have a bandwidth of about 1 eV and 
are well separated from other bands.  These bands contain 3 electrons per O$_2^-$ ion,
so the occupations of the $\pi^*_x$ and $\pi^*_y$ (hereafter $|x\rangle \equiv \pi_x^*$ and 
$|y\rangle \equiv \pi_y^*$) orbitals are frustrated and likely related
to the structural transitions, as well as the Jahn-Teller distortion that tilts the O$_2^-$ ions.   
Since the relevant bands are so narrow and the system is Mott insulating, one 
might consider the use of the LDA+U method \cite{LDAUreview}, but since the typical
implementation of LDA+U in DFT codes uses interactions between onsite atomic orbitals and
the appropriate interactions in MO$_2$ would be between molecular orbitals, a straightforward 
application of LDA+U fails to produce the correct ground state \cite{Winterlik09}.

We used the FPLO code \cite{FPLO} to construct a tight-binding Hamiltonian by projecting Wannier functions using 
symmetry projected orbitals \cite{Wei-PRL,ylvisaker:075104} corresponding to the two O$_2$ molecular $\pi^*$ orbitals 
in the primitve cell in a paramagnetic LDA calculation.  This allows tight-binding hopping parameters to be calculated 
directly.  We consider only the four most relevant hoppings, two for nearest neighbors ($t_{xx} = \langle \mathbf{0}x|
H|\mathbf{R}_1x\rangle$ and $t_{xy} = \langle \mathbf{0}x|H|\mathbf{R}_1y\rangle$, where $\mathbf{R}_1 = 
\tfrac12(a,a,c)$) and two for second neighbors in the plane ($t_\sigma = \langle\mathbf{0}x|H|\mathbf{R}_{2x}x\rangle$ 
and $t_\pi= \langle\mathbf{0}x|H|\mathbf{R}_{2y}x\rangle$ with $R_{2x} = a\hat{x}$ and $R_{2y} = a\hat{y}$).  The LDA 
spin-unpolarized band structure and our tight binding band structure with four parameters are shown in Fig. 
\ref{fig:tb}a.  A schematic for the hopping channels used in the tight-binding model is shown in Fig.  \ref{fig:tb}c 
with the numerical values of the hoppings.  Note that the largest hopping is $t_{xy}$, which represents nearest 
neighbor hopping from $|x\rangle$ to $|y\rangle$ orbitals.  Second neighbor hoppings from $|x\rangle$ to $|y\rangle$ 
are forbidden by symmetry.  There are 8 first neighbors with $t_{xy}$ hopping, compared to 2 each for $t_\sigma$ and 
$t_\pi$, and since $8|t_{xy}| >> 2|t_\sigma| + 2|t_\pi|$, this would suggest that the strongest AFM coupling is 
between nearest neighbors.  Previous neutron diffraction studies \cite{Smith66} on KO$_2$ show magnetic ordering
consistent with nearest neighbor spin ordering.

The non-interacting Hamiltonian is 2x2, with values
\begin{eqnarray}
		H_{xx}^0 & = & -8t_{xx} \gamma(\mathbf{k}) - 2t_\sigma \cos k_x - 2 t_\pi \cos k_y 	\\
		H_{yy}^0 & = & -8t_{xx} \gamma(\mathbf{k}) - 2t_\pi \cos k_x  - 2 t_\sigma \cos k_y 	\\
		H_{xy}^0 & = & -4t_{xy} \cos \left(\tfrac{1}{2} k_z\right)\sin\left(\tfrac12 k_x\right)\sin\left(\tfrac12 k_y\right)
		\label{eq:Hxy}
\end{eqnarray}
where we define the nearest neighbor structure factor
\begin{equation}
    \gamma(\mathbf{k}) = \cos \left(\tfrac12 k_x\right) \cos\left(\tfrac12 k_y\right) \cos\left(\tfrac12 k_z\right).
\end{equation}
The interacting Hamiltonian is
\begin{equation}
	H = \sum_{i\alpha,j\beta} t_{i\alpha,j\beta} c_{i\alpha}^\dagger c_{j\beta} 
			+ \tfrac12 U \sum_{i,\alpha,\beta} n_{i\alpha} n_{i\beta}
\end{equation}
where $i,j$ run over sites and $\alpha,\beta \in [x,y]$.  We apply second-order perturbation theory to $H$
to get an effective Kugel-Khomskii (KK) \cite{KugelKhomskii} type Hamiltonian $H_{KK} = H_{KK}^{nn} + H_{KK}^{nnn}$ for
nearest and next nearest neighbor interactions,
\begin{eqnarray}
	H_{KK}^{nn} &=&\sum_{\langle i,j \rangle} {\Bigl \{}\tfrac12(J_{xy} + J_{xx})\left(\SdotS - \tfrac34\right)   \nonumber \\
	  &+& {\Bigl[} \tfrac14 J_{xy}(\tau_i^+ \tau_j^+ + \tau_i^- \tau_j^-) \nonumber \\
	   &&+ \tfrac14J_{xx} (\tau_i^+ \tau_j^- + \tau_i^- \tau_j^+) \nonumber \\
 	  &+& \tfrac12(J_{xx} - J_{xy})\tau_i^z \tau_j^z 	{\Bigr]} \left(\SdotS + \tfrac 14 \right)  
	  {\Bigr \}} \\
	H_{KK}^{nnn} &=& \sum_{[ i,j ]} \Bigl\{ \tfrac14 J_s \Bigl[ \SdotS -\tfrac34 +  \tau_i^z \tau_j^z \left( \SdotS + \tfrac14 \right) \Bigr] \nonumber \\
	&+& \tfrac14{J_d}(\tau_i^z + \tau_j^z)\left(\SdotS-\tfrac14 \right)  \nonumber \\
	&+& J_I(\tau_i^+ \tau_j^- + \tau_i^- \tau_j^+) \left(\SdotS + \tfrac14\right)	
	  {\Bigr \}}
\end{eqnarray}
in terms of parameters $J_{xy} = 4t_{xy}^2/U$, $J_{xx} = 4t_{xx}^2/U$, $J_\sigma = 4t_{\sigma}^2/U$, 
$J_\pi = 4t_{\pi}^2/U$, $J_s = J_\sigma + J_\pi$ where terms with a single $\tau^\pm$ operators have been neglected.  
The parameter $J_d = \pm(J_\sigma - J_\pi)$ depends on the direction of the bond between $i$ and $j$. The $\tau$ 
operators can be represented as Pauli matrices that operate in orbital space. For numerical results, we select 
$U = 3$ eV, which is equivalent to $U_\mathrm{eff} = U-J$ for intraorbital $U$ and Hund's exchange $J$ found previously
\cite{Solovyev08} for the O$_2$ $\pi^*$ orbitals in KO$_2$ via the constrained LDA method \cite{CLDA}.  This gives 
$J_{xx} = 1.54$ meV, $J_{xy} = 4.8$ meV, $J_\sigma = 3.89$ meV, and $J_\pi = 0.34$ meV.

Unlike previous studies \cite{Khaliullin97, Feiner97, HarrisLandau, HarrisUnusualSym} on the KK Hamiltonian, which
contained only a single parameter $t$ for both intraorbital and interorbital hopping, we consider here a KK-type
Hamiltonian where the hoppings between orbitals channels are significantly different than hoppings within an channel, as
shown in Fig. \ref{fig:tb}.  One previous study \cite{HarrisUnusualSym} concluded that due to unusual symmetries present,
the KK Hamiltonian could not describe the observed order and gapped excitations.  In the present work, the broken 
symmetry of the hoppings in this model should avoid these difficulties.  These hoppings will lead to the orbital 
ground state having a significant impact on the spin wave dispersion, which should be measurable in experiment.

\begin{figure}[t]
\begin{centering}
\includegraphics[width=\columnwidth]{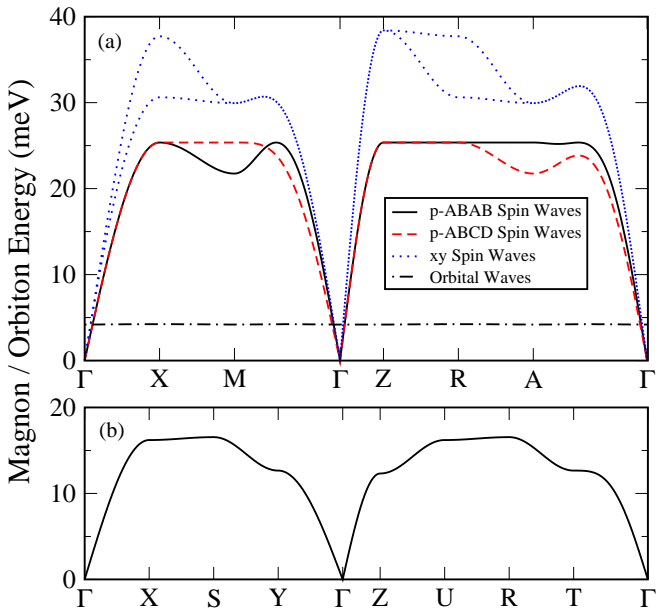}
\caption{(Color online) (a) Spin wave spectrum for RbO$_2$ for p-type ordering for both stackings as described in text (black solid, ABAB stacking and red dashed, ABCD stacking), and spin wave spectrum for xy ordering (blue dotted).  Also, the nearly dispersionless orbital spectrum is shown for p-type orderings (black dash-dot line). (b) Spin wave spectrum for ferroorbital ordering, which is orthorhombic symmetry.}
\label{fig:magnon}
\end{centering}
\end{figure}

To proceed with spin/orbital wave theory a reference ground state for both the spin and orbital systems must be 
chosen.  From here on, we will consider the electronic structure from the hole perspective, so that there is a single 
hole per site.  We restrict our magnetic order to be antiferromagnetic between with nearest neighbors.

The ground state orbital ordering for this model is where antiferroorbital (AFO) ordering occurs in planes so that a 
given site with, say, $|x\rangle$ occupied would have second neighbors (first neighbors in the plane) with $|y\rangle$
occupied.  We refer to this ordering as p-type.  This ordering frustrates the first neighbor orbitals, so alternate
stackings of the planes will be very close in energy and may be degenerate.  For these orderings, the spin wave 
dispersion $\omega_\mathbf{k}^{p}$ and orbital wave dispersion $\nu_\mathbf{k}^{p}$ are given by
\begin{eqnarray}
	\omega_\mathbf{k}^{p} &=& \sqrt{(8\bar J)^2 - \left[4 J_{xy} \gamma_m(\mathbf{k}) + 4 J_{xx} \gamma_n(\mathbf{k})\right]^2} \\
	\nu_\mathbf{k}^{p} &=& \sqrt{ J_s^2 - \left(\tfrac12 J_I \gamma_2(\mathbf{k})\right)^2 } 
\end{eqnarray}
where $J_I = \sqrt{J_\sigma J_\pi}$ and $\gamma_2(\mathbf{k}) = \tfrac12 (\cos k_x + \cos k_y)$.
where the effect of different stackings is contained in the structure factors $\gamma_m(\mathbf{k})$ 
and $\gamma_n(\mathbf{k})$.
First we consider the case of ABAB stacking, where any given site has the same orbital occupation as the sites 
at displacements $(0,0,c)$ and $(0,0,-c)$ from it.  This results in structure factors given by 
\begin{eqnarray}
	\gamma_m(\mathbf{k})^\mathrm{ABAB} &=& \cos\left( \tfrac12 (k_x + k_y) \right) \cos (\tfrac12 k_z) \nonumber \\
	\gamma_n(\mathbf{k})^\mathrm{ABAB} &=& \cos\left( \tfrac12 (k_x - k_y) \right) \cos (\tfrac12 k_z).
\end{eqnarray}
An alternate stacking, ABCD, where each site has the opposite orbital occupation as the sites above and below in 
the $\hat{z}$ direction, 
has structure factors of
\begin{eqnarray}
	\gamma_m(\mathbf{k})^\mathrm{ABCD} &=& \cos \tfrac{k_x}2 \cos \tfrac{k_y}2 \cos \tfrac{k_z}2 + \nonumber \\
			&& i \sin \tfrac{k_x}2 \sin \tfrac{k_y}2 \sin \tfrac{k_z}2 \nonumber \\
	\gamma_n(\mathbf{k})^\mathrm{ABCD} &=& \cos \tfrac{k_x}2 \cos \tfrac{k_y}2 \cos \tfrac{k_z}2 - \nonumber \\
			&& i \sin \tfrac{k_x}2 \sin \tfrac{k_y}2 \sin \tfrac{k_z}2.
\end{eqnarray}
The dispersions are depicted in Fig. \ref{fig:magnon}a.  

An alternate orbital ordering we consider is where the ordering within planes are ferroorbital (FO) ordering.  This
ordering is not stable
to orbital fluctuations within our KK model, however it may be stabilized in RbO$_2$ by effects not considered here, 
such as the Jahn-Teller distortion that causes the canting of the O$_2$ molecules.  This ordering is of particular
interest because if the stacking of planes is AFO, then the nearest neighbor spin exchange is maximized.  With this
orbital ordering (hereafter referred to as xy ordering),the spin sublattices have different dispersions due to the
swapping of $J_\sigma$ and $J_\pi$ in each plane.  The 
Kugel-Khomskii Hamiltonian yields a spin wave dispersion of
\begin{eqnarray}
	\omega_\mathbf{k}^{xy} &=& \sqrt{A_m^2 - (8J_{xy}\gamma(\mathbf{k}))^2} \pm A_d 
\end{eqnarray}
with $A_m = 8 J_{xy} - J_s\left[2 - \cos(k_x) - \cos(k_y)\right]$ and $A_d = (J_\sigma - J_\pi) \left[\cos(k_x) - \cos(k_y)\right]$, 
and is shown in Fig. \ref{fig:magnon}a.  

Thus far, the orderings considered all preserve tetragonal symmetry, but orthorhombic and monoclinic low temperature
phases of RbO$_2$ exist.  The final orbital ordering we consider is FO ordering for every site in the crystal, hereafter
called xx ordering, which is orthorhombic symmetry.  This has spin-wave dispersion
$\omega_\mathbf{k}^{xx} = \sqrt{A_\mathbf{k}^2 - B_\mathbf{k}^2}$ where
$A_\mathbf{k} = 8 J_{xx} + \tfrac12(J_s - J_\sigma \cos k_x - J_\pi\cos k_y)$ and $B_\mathbf{k} = 8 J_{xx} \gamma(\mathbf{k})$.
Again, this configuration is not stable with respect to orbital fluctuations, but it may be stabilized by effects not
considered here, which a study of KO$_2$ indicates is the case \cite{Kim09}.  The magnon dispersions is shown in Fig.
\ref{fig:magnon}b.

\begin{table}[th]
\begin{center}
\begin{tabular}{c|cccc}
\multicolumn{1}{c}{}&\multicolumn{4}{c}{\bf Orbital Order}  \\
\hline
\hline
\multicolumn{1}{c}{}&	xx      &	xy	     & p-ABAB    & p-ABCD    \\	\hline \hline
$E_0$      & -12.680  & -12.680  & -14.795   & -14.795   \\
$E_{S}$    &  -0.411  &  -3.266  &  -2.333   &  -2.288   \\
$E_{O}$    &   0      &   0      &  -0.010   &  -0.010   \\
$E_{tot}$  & -13.091  & -15.946  & -17.138   & -17.093   \\
\hline\hline
\end{tabular}
\end{center}
\caption{Energies contributing to the ground state energy in meV.  $E_0$ is the classical ground state energy, $E_{S}$($E_{O}$) is the quantum correction from spin(orbital)-wave theory.}
\label{tbl:energy}
\end{table}

The ground state energies for these four orbital orderings are listed in table \ref{tbl:energy},
where we find that the ABAB stacking of planar orbital ordering is the lowest energy.  For an
average nearest neighbor spin exchange of $\bar J = 3.17$ meV (in p-type orderings), high
temperature series expansion \cite{Oitmaa04} would predict $T_N \approx 1.4 \bar J = 51 K$, much
higher than the observed $T_N = 15 K$.  The frustrated exchanges within planes would reduce this
value somewhat, but not enough to give a prediction reasonably close to the experimental transition.  

As expected, the two stackings examined for the planar orbital ordering are very nearly degenerate,
with only quantum fluctuations in the spin waves breaking the degeneracy at this level of
approximation.  It is rather clear that due to the strong asymmetry between hopping within an
orbital channel and hopping between orbital channels, the orbital ground state significantly impacts
the spin wave spectrum and could be inferred from a measurement of the low temperature magnons.  The
planar orbital orderings don't significantly impact the spin wave dispersion, so even if the
stacking is disordered magnon excitations should be coherent.  Energetically the next state above p-
type ordering is the xy ordering, which is $\sim$1.2 meV $\approx$ 14 K higher in energy than p-type
ordering.  Above this temperature orbital domains should proliferate.  The strong modulation of 
exchange constants at the domain boundaries should nucleate magnetic domains \cite{Mazin09}, leading 
to the low observed N\'eel temperature of RbO$_2$ of 15 K.

Orbitons are quite difficult to measure experimentally.  They do not couple directly to neutrons, the 
standard measurement technique for magnons.  Recently orbitons have been measured in titanates via 
X-rays \cite{orbiton}, but the inference of the orbital dispersion is very indirect.  The measured 
dispersion is negligible, suggesting \cite{orbiton} that X-rays modulate bonds resulting in 
a much bigger scattering from two-orbitons than from single orbitons. 
It may be that in narrow band molecular oxide systems the X-rays will create single orbiton excitations 
with all bond modulation inside a unit cell.  
If that is the case, the X-ray measured dispersion in RbO$_2$ should be clearly observable, if a similar
experiment can be done.  Neglected here is the Jahn-Teller effect which rotates the O$_2$ molecules
and breaks the orbital degeneracy, which will select a particular ground state orbital ordering.

We have examined independent magnon and orbiton excitations in RbO$_2$ within spin/orbital wave theory, 
finding considerable coupling between the easily measured magnon excitations and the difficult to measure 
orbital ground state.  This strong coupling arises from the large anisotropy in the hopping parameters 
in the O$_2$ $\pi^*$ bands.  This indicates that MO$_2$ materials are attractive for studying the 
interplay between orbital ordering and $p$-electron magnetism.  E. R. Y. and W. E. P were supported by 
DOE SciDAC Grant No. DE-FC02-06ER25794.  The authors would like to thank J. Kune\v{s}, R. T. Scalettar, 
and C. Felser for stimulating conversation.

\bibliographystyle{apsrev}
\bibliography{rbo2}

\end{document}